# Wi-Motion: A Robust Human Activity Recognition Using WiFi Signals

Heju Li, Xukai Chen, Haohua Du, Xin He, Jianwei Qian, Peng-Jun Wan and Panlong Yang

**Abstract**

Recent research has shown that human motions and positions can be recognized through WiFi signals. The key intuition is that different motions and positions introduce different multi-path distortions in WiFi signals and generate different patterns in the time-series of channel state information (CSI). In this paper, we propose Wi-Motion, a WiFi-based human activities recognition system. Unlike existing systems, Wi-Motion adopts the amplitude and phase information extracted from the CSI sequence to construct the classifiers respectively, and combines the results using a combination strategy based on posterior probability. As the simulation results shows, Wi-Motion can recognize six human activities with the mean accuracy of 98.4%.

**Index Terms**

WiFi Signals, Human Activity Recognition, Support Vector Machine (SVM), Posterior Probability Combination

## I. Introduction

As experiments show, the machine-centric computing model is shifting toward a people-centric computing model [1], [2], where it is critical to precisely sense and recogniz human activities. Conventional methods for recognizing human activities can be categorized into three groups: vision-based approaches, low-cost radar-based approaches and wearable sensor-based approaches. However, all of these conventional approaches have some limitations. Vision-based approaches are susceptible to lighting conditions and obstacles. At the same time, the camera has a dead angle where it may breach human privacy, resulting in the perception within only a certain range of line of sight. Low-cost radar-based systems have limited operation ranges of just tens of centimeters. Wearable sensor-based solutions although can achieve fine-grained behavioral awareness, but high cost and restriction on real-time nature, make it not practical in some applications(e.g. rescue applications).

In recent years, with wide deployment of WiFi hotspots and rapid development of WiFi-based indoor Sensing technology, several WiFi-based approaches have been proposed to recognize human activity. We can distinguish different human activities by detecting and analyzing different multi-path distortions in WiFi signals, which

This research is supported in part by the National Natural Science Foundation of China (NSFC) under Grant 61702011, and in part by the National Science Foundation of USA under Grant CNS-1526638. *Corresponding author*: Xin He.

H. Li and X. Chen are with the School of Computer and Information, Anhui Normal University, No. 189 South Jiuhua Road, Wuhu, Anhui, China 241002 (e-mail:{762699132, 373456543}@qq.com).

H. Du, J. Qian and P-J. Wan are with the Department of Computer Science, Illinois Institute of Technology, Chicago, IL 60616 USA (e-mail: {hdu4, jqian15}@hawk.iit.edu, wan@cs.iit.edu).

X. He and P. Yang are with the School of Computer Science and Technology, University of Science and Technology of China, Hefei, Anhui, China (e-mail: {xhe076, plyang}@ustc.edu.cn).

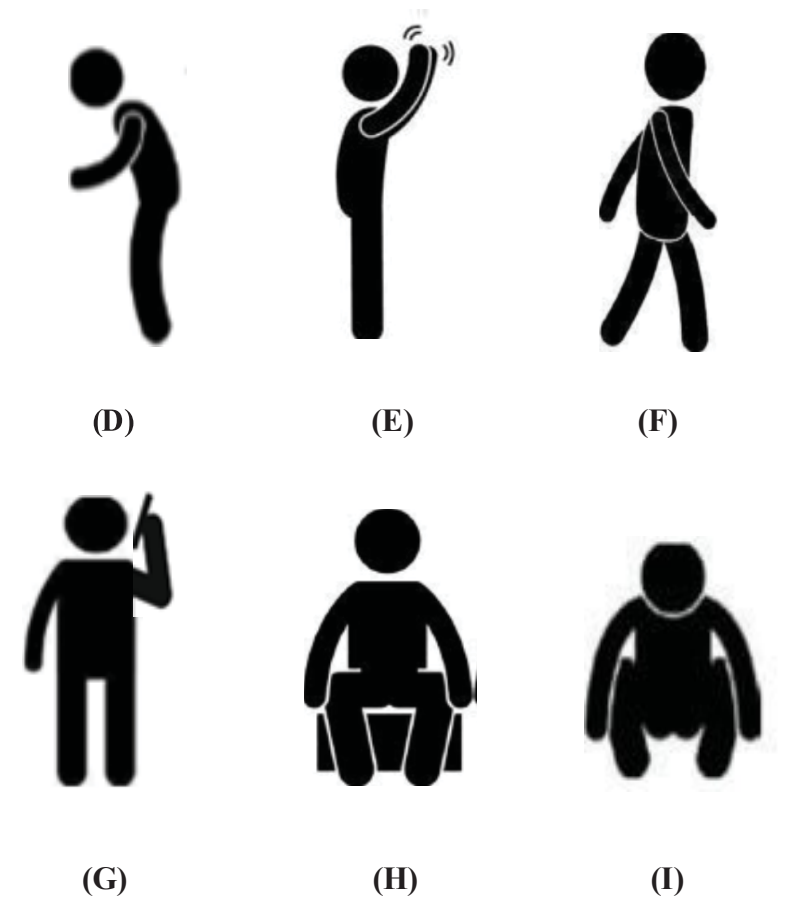

Fig. 1. Six human activities recognized by Wimotion. (a) Bend. (b) Hand Clap. (c) Walk. (d) Phone Call. (e) Sit Down. (f) Squat.

generate a unique pattern in the time-series of channel state information values. Each CSI depicts the amplitude and phase of a subcarrier.

$$H(f_k) = ||H(f_k)||e^{j\sin(\angle H)}. \tag{1}$$

where $H(f_k)$ indicates the CSI at the subcarrier with central frequency of $f_k$ , and $||H(f_k)||$ and $\angle H$ denote its amplitude and phase, respectively. A group of CSIs $H(f)$ of $K$=30 subcarriers are exported to upper layers.

$$H(f) = [H(f_1), H(f_2), ..., H(f_K)]. \tag{2}$$

Some commercial WiFi devices (e.g. IWL5300 wireless network card and sora simulation platform) provide us a fine-grained CSI in time-series. Because of the high data rate provided by these modern commercial WiFi devices, we can get enough samples of CSI measurements within the duration of human activities. Take advantage of these systems, researchers are more comfortable using CSI to recognize various types of human activities. E-Eyes [3] is a smart home human behavior perception system proposed by Wang *et. al.* in 2014. By collecting CSI information on commercial WiFi devices, E-Eyes senses and recognizes 11 actions and 9 cross-room walks in a single environment. WIAR [4] is a CSI-based human behavior analysis and monitoring system proposed by Guo *et. al.*, which establishes a WiFi-based activity dataset, as a benchmark to evaluate the performance of existing activity recognition systems. These systems follow the general architecture of machine learning-based systems and generally have four stages: data collection, noise removal, feature extraction, and classifcation.

In this paper, we propose a WiFi-based human activity recognition system, namely Wi-Motion, using the open source datasets built by Guo *et. al.* [4] with IWL5300 wireless network card, and choose six most common human activities in daily home life, as shown in Fig. 1. To summarize, the contributions of this paper are shown as follows:

- Unlike most human body recognition systems, which only process amplitude information, we furthermore extract the phase information in the CSI sequence and mathematically eliminate its random offset. Then we leverage several signals processing methods to obtain the high-quality dataset.

- We use different methods to extract the feature of the phase and amplitude information separately, designing different classifier with different methods. For phase feature, we chose the appropriate kernel function to build the SVM classifier after many experiments. We perform a unique classification algorithm by combining DTW algorithm and SVM model,adjusting the relationship between different amplitude feature vectors with different dimensions , construct a new SVM kernel function.
- After getting the recognition results at each classifier, we combined prediction results based on output posterior probability of two classifiers. In order to verify the effectiveness of the combine algorithm, we design a large number of comparative experiments. According to the experiment results, the solutions we proposed increase recognition accuracy clearly.

In the rest of this paper, we will present the related work in Section 2. Then we elaborate the design details and classification algorithm of Wi-Motion in Section 3 and 4. We present the implementation and evaluation in Section 5 and finally conclude our work in Section 6.

## II. Related Work

### *A. Hardware-based Methods*

WiSee [5] used USRP as wireless devices and utilizes communication on a 10 MHz channel at 5 GHz. This implementation can recognize nine actions by extracting the doppler shift of human motion from the WiFi signal as a feature, with an accuracy rate of 94%. Adib *et al.* designed WiTrack and WiTrack2.0 that apply specially designed carrier wave radio to track human movements behind a wall [6].

### *B. RSS-based Methods*

As early as 2000, Bahl *et al.* proposed Radar [7], which is a system for indoor localization, based on received signal strength (RSS) . This is the first time where WiFi signals have been used for perception. Sigg *et al.* used USRPs as specialized hardware devices to capture RSS values from WiFi signals [8], [9]. They utilized RSS values of WiFi signals to recognize four activities including lying down, crawling, standing and walking and achieved over 80% recognition accuracy for these four activities. Abdelnasser *et al.* proposed WiGest [10], a gesture recognition system based on universal WiFi received signal strength. Based on existing devices, WiGest performs gesture recognition by analyzing the rising and falling edges of RSS signal changes. The accuracy can reach 87.5% in the case of a single access point, and it can increase to 96% when there are three access points. However, since RSS only provide coarse-grained information about channel variations and do not offer fine-grained information about small scale fading and multi-path effects caused by micro-movements, it is often affected by multipath effects and noise signals.

### *C. CSI-based Methods*

Compared with RSS, CSI provides not only fine-grained channel status information, but information about small scale fading and multi-path effects caused by micro-movement. WiHear [11] further interpreted the transmitted signal by directing it to the human mouth and analyzing the changes in the mouth shape by reflecting the signal. The pronunciation was represented by the mouth type, thereby implementing WiFi-based lip recognition. WiFinger [12] extracted the fixed pattern of gesture signals through principal component analysis (PCA), and used it as a feature to identify gestures with an accuracy of 93%. Ali *et al.* proposed WiKey that uses CSI values obtained from COTS to recognize keystrokes [13]. Zheng *et al.* built a novel non-intrusive smoking detection system, namely Smokey [14], that is able to accurately detect the smoking activities by exploiting the impact of smoking on the CSI of WiFi. Shang *et al.* propose a WiFi signal-based sign language recognition system called WiSign [15]. Different from other systerm, WiSign uses 3 WiFi devices to improve the recognition performance. These splendid research specialized in specific application scenarios which do not contain continuous text input using CSI characteristics. With the inspiration of the above mentioned works, we propose a framework of WiFi-based activity recognition to improve the human activity recognition.

## III. System Design

### *A. System Structure*

In this section, we elaborate the design of Wi-Motion. Wi-Motion is a wireless system that enables commercial WiFi devices to identify people's activity using Orthogonal Frequency Division Multiplexing (OFDM) technology. The system flows of Wi-Motion are illustrated in Fig. 2. Firstly, a signal containing human activity information is acquired from a specific receiving device. Secondly, the collected signal, which is separated into amplitude and phase information, should be respectively subjected to preprocessing such as filtering and linear transformation to reduce noise and obtain useful information. Since each CSI information contains 30 subcarriers, this will result in too many dimensions of the data, causing the complexity of the system to become high. Additionally, some subcarriers may be more sensitive to human motions. Utilizing all the subcarriers is therefore not wise because the intrinsic noise on some subcarriers can be too serious to conceal the meaningful information about motions if the subcarriers are sensitive to noise but insensitive to human motions. Therefore, it is essential to reduce the dimensionality of the filtered data.

After dimensionality reduction, we extract useful features from the processed amplitude and phase information respectively. Since CSI waveforms of different activities differ on some features, so we can extracte suitable feature in both amplitude and phase information , which can represent the relationship between the time-series of CSI and different human activities, as a basis for classification. In the classification stage, we randomly select parts of feature vectors, using SVM algorithm to build two classifiers leveraging amplitude and phase information, respectively. When unknow activity enters, according to the prediction results of both two classifiers, Wi-Motion perform a merge method based on posterior probability to produce the final recognition.

### *B. Phase Information Preprocessing*

*1) Phase Analysis :* As discussed in section 1, CSI measurements provide the phase information of each subcarrier. The separated phase information $\hat{\varphi}_i$ for the $i^{th}$ subcarrier can be expressed as:

$$\hat{\varphi}_i = \varphi_i - 2\pi \frac{k_i}{N} \delta + \beta + Z. \tag{3}$$

where $\varphi_i$ denotes the true phase, $\delta$ is the timing offset at the receiver, which causes phase error expressed as the middle term, $\beta$ means an unknown phase offset, and $Z$ indicates some measurement noise. $k_i$ signifys the subcarrier index (ranging from −28 to 28 in IEEE 802.11n) of the $i^{th}$ subcarrier and $N$ represents the FFT size (which sets as 64 in IEEE 802.11 a/g/n). Due to the unknowns listed above, it is impracticable to obtain the true phase shifts with solely commercial Wi-Fi devices.

*2) Phase Calibration :* To mitigate the effects of random noise, we execute a linear transformation on the raw phases, as recommended in [16]. The key thoughts is to remove $\delta$ and $\beta$ by considering phase across the entire frequency band. Firstly, we define two intermediate items $a$ and $b$ as follows:

$$a = \frac{\hat{\varphi}_n - \hat{\varphi}_1}{k_n - k_1} = \frac{\varphi_n - \varphi_1}{k_n - k_1} - 2\pi\frac{\delta}{N}. \tag{4}$$

$$b = \frac{1}{n}\sum_{j=1}^{n}\hat{\varphi}_j = \frac{1}{n}\sum_{j=1}^{n}\varphi_j - \frac{2\pi\delta}{nN}\sum_{j=1}^{n}k_j + \beta. \tag{5}$$

where a and b indicate the slope of phase and the offset across the entire frequency band, respectively. If the subcarrier frequency is symmetric, which means $\sum_{j=1}^{n} k_j = 0$, $b$ can be expressed as $b = \frac{1}{n}\sum_{j=1}^{n}\varphi_j + \beta$. Subtracting the linear term $ak_i + b$ from the raw phase $\hat{\varphi}_i - ak_i$ in Equation 4, we can get a linear combination of true phases, denoted as $\tilde{\varphi}_i$ , from which the random phase offsets have been eliminated (omitting the small measurement noise $Z$ ).

$$\tilde{\varphi}_i = \hat{\varphi}_i - ak_i - b = \varphi_i - \frac{\varphi_n - \varphi_1}{k_n - k_1}k_i - \frac{1}{n}\sum_{j=1}^{n}\varphi_j. \tag{6}$$

Although the above equation 6 can be used for calibrating phase information, the raw phase is folded due to the recurrence characteristic of phase, which requires us to map the raw phase into the true value. Fig. 3 shows the raw phase values of CSI for the three antennas at the receiver. What we can clearly see is that the raw phase of each of the three antennas is folded with the increase of subcarrier order and the range of the phase is [-π π]. To obtain the true phase, the folded phase can be recovered by subtracting multiple 2π.

**Algorithm 1** Phase Calibration

**Input:** raw phase $M_P = \hat{\varphi}_i$ of 30 subcarriers;
**Output:** transformed phase $C_P = \tilde{\varphi}_i$ of 30 subcarriers;

```
Set T_P as a vector as the same size of M_P;
Set k as a vector from -28 to 28;
Set diff = 0;
Set η = π;
Set T_P(1)=M_P(1);
for i = 2 to 30 do
    if M_P(i) - M_P(i-1) > η then
        diff = diff + 1;
    end if
    T_P(i) = M_P(i) - diff*2*π;
end for
Compute a = (T_P(30) - T_P(1)) / (k(30) - k(1));
Compute b = sum{T_P}/30;
for i = 1 to 30 do
    C_P(i) = T_P(i) - a*k(i) - b;
end for
```

Thus, we perform a phase calibration algorithm in Algorithm 1 proposed by Wang *et al* [17]. Fig. 4 shows the transformed phase values for three different antennas. It is noticed that the range of the transformed phase becomes much smaller than the raw phase for three antennas. Fig. 5 makes a comparison of unprocessed raw phase and transformed phase information of the first subcarrier of a squat sample. As can be seen, the phase without further calibration distribute extremely randomly. But after calibration, it behave relatively stably as expected.

### *C. Amplitude Information Preprocessing*

*1) Noise Remove Algorithm:* The raw amplitude waveform we separate from raw CSI measurements is usually not reliable enough to be used for feature extraction because of the noise, which can be from environmental changes, radio signal interference, etc. In our system, we further introduces weighted moving average (WMA) method to the raw amplitude waveform. $\{AMP_{1,1}, ..., AMP_{t,1}\}$ denotes the amplitude value sequence of first subcarrier in the time period $t$ ,the expression of amplitude series is shown as follows.

$$AMP_NEW_{t,1} = \frac{(m \times AMP_{t,1} + ... + 1 \times AMP_{t-m-1,1})}{m + (m-1) + ... + 1}. \tag{7}$$

where *AMP_NEW* indicates the averaged new amplitude, the value of $m$ decides in what degree the current value is related to historical records. In this paper, we set $m$=10. Fig. 6 shows the original waveform of the first subcarrier of squat and the waveform after WMA filtering. Comparison shows that WMA filtering can remove most of the noise, which makes waveform smoother.

*2) Dimensionality Reduction:* The IWL5300 provides 802.11*n* channel state information in a format that reports the channel matrices for 30 subcarrier groups. At each subcarrier, the fine-grained CSI describes how a signal propagates from the transmitter to the receiver with the combined influence of, for example, scattering, fading, and power decay with distance. After noise remove, we can get a relatively accurate amplitude matrix of each activity sample. However, if all subcarriers are used to perform the following operations, it will definitely cause the complexity of the system to become higher. On the other hand, as show in Fig. 7, we notive clearly that different subcarriers have different sensitivities for same activity. When data is affected by noise, some subcarriers that are very sensitive to noise but show very low sensitivity to human activity will unpredictable hinder the work behind. Therefore, reducing the data dimension and eliminating these non-significant subcarriers is very important. In this paper, we leverage PCA algorithm to reduce the dimensions of the CSI sequence and eliminate redundant information remaining in data sequence. Based on our experiment results, we finally choose the first principal component waveform for subsequent operations, which was showed in Fig. 8.

## IV. Classification

### *A. Feature Extraction*

*1) Amplitude Feature Extraction:* Discrete wavelet transform (DWT) can analyze signals on multiple frequency scales and has better extraction ability for local features. Considering that the speed of movement of different parts of the body is different, direct extraction will lose a lot of detail. Through the wavelet transform, the wavelet coefficients of each frequency band are obtained. For the first principal component obtained from amplitude waveform after PCA processing, it maintains most features of the raw amplitude waveform, so the features corresponding to each frequency band can be extracted. Firstly, we perform DWT processing on the extracted amplitude waveform based on the first-order Daubechies wavelet, where the decomposition layer number is 3. In Wi-Motion, several wavelet families have been tested such as Daubechies, Coiflets, Symlets. Due to the classification performance, the Daubechies D1 coefficient wavelet family is selected. Then, the approximate coefficient of the last layer is taken out, and the normalized coefficient sequence is used as the feature vector. After feature extraction, the contour information of the amplitude waveform is preserved in the feature vector, and the noise is suppressed as the detail coefficients are discarded. The complete binary tree of the DWT process can be shown in Fig. 9.

*2) Phase Feature Extraction:* For the phase information matrix, it also includes phase information of 30 subcarriers. In this paper, we use singular value decomposition (SVD) to simplify the CSI phase difference matrix of the first receive antenna and the second receive antenna. SVD is a method with obvious physical meaning. It can represent a more complex matrix by multiplying smaller and simpler sub-matrices, which describe the important characteristics of the raw matrix. Based on our experiment results, the top 5 singular values of the SVD matrix are more useful for classifcation.

### *B. Classifier Training*

We select a high effective SVM classification to recognise six activities according to the performance of existing works. As is known to all, the choice of kernel function plays a key role in the performance of classical SVM. For example, a Gaussian kernel function that is simple in form and widely used.

$$K(x, x_i) = \exp[-\mid x - x_i \mid^2 / 2\delta^2]. \tag{8}$$

where vector $x_i$ represents the center of the kernel function, and $\mid x - x_i \mid^2$ represents the Euclidean distance of any vector $x$ to the center of the kernel function.

For our extracted amplitude features, we find that the feature vectors of human activities may not share the same length, so the traditional SVM algorithm requiring the dimension of the feature vector to be consistent cannot be applied in classifying our amplitude features. In this situation, we use dynamic time warping (DTW) to calculate the distances among feature vectors. In contrast to Euclidean distance, DTW offers intuitive distance between two waveform and can be resilient to signal distortion or shift. DTW distance is the Euclidean distance of the optimal warping path between two waveforms calculated under boundary conditions and local path constraints [18]. The aim of DTW is to compare two timedependent series $X = (x_1, x_2, ..., x_n)$ of length $n \in N^+$ and $Y = (y_1, y_2, ..., y_m)$ of length $m \in N^+$. These sequences can be discrete signals (time series) or, more generally, feature sequences sampled at equidistant points in time. Therefore, we use DTW distance to replace the Euclidean distance in the Gaussian kernel function to construct a new kernel function.

$$K(x, x_i) = \exp[-DTW(x, x_i)^2 / 2\delta^2]. \tag{9}$$

Finally, we classify our amplitude feature vectors using the support vector machine with the kernel function defined in the equation 9. For our extracted phase features, we don't have to worry about the above problem, where the dimensions are inconsistent, because we chose the same quantity of singular value. We tested various kernel functions of the SVM, such as linear kernels, Gaussian kernels, and polynomial kernels, etc. According to classification performance, we choose a Gaussian kernel whose classification performance is much larger than other kernel as the final kernel function of phase SVM model.

*C. Prediction*

In our experiments, we collect the CSI waveforms of six different activities (“bend”, “hand clap”, “walk”, “phone call”, “squat” and “sit down”) that are defined in Wiar to test our two classifiers. Results are shown in the Fig. 10. We find that for some activity, like “hand clap”, we can use the kernel-based SVM model with phase feature to perfectly classify them. However, for some other activities like “bend”, the classification effect of the phase classifier is not satisfactory. The same situation occurs on the amplitude classifier, where “walk” and “squat” can be classified very well, but other activity like “hand clap” can not be well recognized. In response to this situation, we propose combine the prediction results properly on two classifiers. Traditional result combination algorithms, such as boosting algorithm, multiple decision method, etc. face higher complexity and limitations for more than three classifier. In our experiment, there are only two classifiers, which lead to that the traditional combination algorithms not suitable for the challenges we are facing. In WiSign [15], Shang *et al.* propose a weighted voting on two laptop and get the final prediction result, where they combine two prediction vectors of classifiers on two laptops instead of choosing the result with the highest confidence on one laptop. In the context of Wi-motion, we want to combine the result of two classifiers. Having understood the similarity of two problem, we propose a combination strategy based on output posterior probability of two classifiers, where the classification result of each classifier is given in the form of posterior probability which represents the membership of the sample for each category. According to the method proposed by Platt [19] in a simple two-class problem, the SVM standard output value is mapped to [0,1] using the Sigmoid function to obtain the SVM posterior probability.

$$P(y = 1 | f(x)) = \frac{1}{1 + \exp(Af(x) + B)}. \tag{10}$$

where $P(y = 1| f(x))$ indicates the probability that the sample under the condition of standard output value $f(x)$ is the target class. $A$ and $B$ are parameters that need to be optimized, which can be obtained by using the training set for maximum likelihood estimation. That is, the target model can be expressed as fellowing formula.

$$\min_{t} F(A, B) = -\sum_{i=1}^{n} [t_i \ln(p_i) + (1 - t_i) \ln(1 - p_i)]. \tag{11}$$

where

$$t_i = \begin{cases} \frac{n_+ + 1}{n_+ + 2}, & y_i = +1 \\ \frac{1}{n_- + 2}, & y_i = -1 \end{cases} \tag{12}$$

where $(x_i, y_i)$ represents the training sample, $n_+$ and $n_-$ indicate the number where the category is +1 and −1 in the training sample, $p_i = \frac{1}{1+\exp(Af(x_i)+B)}$. Wi-Motion is a six-class task. In our experiments, we extend the two-class probability based SVM to the multi-class in a one-to-one manner, where we need to synthesize 6*(6-1)/2=15 results for each classifier (amplitude and phase).

After the test sample enters, two classifiers respectively predict it and generate a posterior probability vector. Then Wi-Motion add the two vectors with same weight and give the final prediction. For example, assume the prediction vector reported by the two classifiers are (0.1, 0.2, 0.13, 0.78, 0.9, 0.27) and (0.12, 0.34, 0.2, 0.87, 0.14, 0.24), we can see that the first classifier cannot distinguish the fourth and the fifth activity. If we always choose the result with the highest confidence on one classifier, it is going to be a wrong prediction. But if we combine these two prediction vectors , we can get (0.22, 0.54, 0.33, 1.65, 1.04, 0.51). Based on the final combined prediction vector, we can get the correct prediction (fourth activity).

## V. Implementation and Evaluation

### A. Activity Dataset

We select the most common six human activities in the home environment from the dataset constructed by Guo *et al* in WiAR [4], as shown in Fig. 1. In WiAR,they use a commercial TP-Link wireless router as the transmitter operating in IEEE 802.11n AP mode at 2.4GHz. A ThinkPad 400 laptop with three antennae running Ubuntu 10.04 is used as a receiver, which is equipped with off-theshelf Intel 5300 card and a modified firmware. During the process of receiving WiFi signals, the receiver pings 30 pkts/s from the router and records the CSI from each packet.

### B. Activity Recognition Accuracy

Fig. 11 shows the mean prediction accuracies of our prediction combination model and classifcation models on each classifier of volunteer 1. We can see that our system can improve recognition performance for all supported activities. Since âĂIJBendâĂİ is unsatisfactory to classify on both amplitude and phase classifier. However, after merging the results of two classifiers, our system have great prediction accuracy of 97%. For âĂIJHand ClapâĂİ, both of the phase classifier and our system have great prediction accuracy of 100%, while the amplitude classifier only has prediction accuracy of 92%. For âĂIJSit DownâĂİ, the prediction accuracies are 95% and 88% on the amplitude and phase classifier, while our system has better result of 98%. The experiment results show that we can get more accurate activity estimation by combining output posterior probability of two classifiers.

### C. Different Number of Training Samples

In this paper, Wi-Motion does not require a large amount of samples, but after our experiments, we notice that the number of training samples has a certain impact on the recognition accuracy. Fig. 12 shows our experimental results. We can find that with the increase of training samples, the classification accuracy of the amplitude and phase classifiers has a certain degree of increase, because the more training samples, the richer the scene, meaning the hyperplane position of the support vector machine is more accurate. Besides, the increase of our combination system is small, because the average recognition rate of our system is already at a very high level. However, what we all know is that the more training samples, the longer time to train the SVM system takes , which inevitably leads to an increase in system complexity. Therefore, we set the number of samples used for training to 110 in our experiments, which minimizing the training time of the sample while ensuring accuracy.

### D. Different Volunteers

In our experiments, we found even for a same activity, the operation range and speed may not be the same since different people tend to have different habits. Thus, to make sure our system can work for different users, we add a new experiment to evaluate the influence generated by different volunteers. We use the trained classifcation model to further evaluate the data collected from the other two volunteers, and the results are shown in Fig. 13. We can find that the prediction performances are still good even if we do not retrain the parameters that need to be set in the model for these two new volunteers. Although the mean prediction accuracies decrease by about 3% and 5% respectively, the recognition accuracy is always at a very high level. What we can expect is that when we retrain new model parameters for these two new volunteers, our system average recognition accuracy must have a obvious increase.

### *E. False Positive and True Positive*

In order to test the recognition performance of Wi-Motion, we further explore the false postive and true positive of each activity supported. We use the same dataset that is used in section 5.1, and the evaluation result is illustrated in Fig. 14 and Fig. 15. It is clear that the false positive of prediction can be improved to about 0.16% by combining two classifiers in most cases. Moreover, the true positive rate generated by the combined model reachs 98.5%, which is significantly higher than that produced by the separate classifier. Therefore, the combination we proposed reduces the recognition mistakes and facilitates the recognition of activity.

## VI. Conclusion

In this paper, we propose a WiFi-based indoor activity recognition system called Wi-Motion. Compared to existing related systems, we adopt both amplitude and phase information constructing classifiers in our system to improve the recognition performance. Moreover, to enhance the robustness of Wi-Motion, the final recognition result of our system is determined by combining prediction results on all classifers based on output posterior probability rather than simply obtaining from single classifer. Experimental results show that our system can get better mean false positive of 0.16% and mean true positive of 98.5%, in addition, improve the recognition accuracy to 98.4% compared with originial implementation that uses only one classifier constructed with amplitude or phase information.

## References

[1] J. Burke,D. Estrin,M. Hansen, âĂIJParticipatory sensing,âĂİ. Proceedings of the 1th Workshop on WorldâĂŘSensorâĂŘWeb:Mobile Device Centric Sensor Networks and Applications.New York,USA,ACM,2006.

[2] D. Estrin,G. Sukhatme,M. Allen,âĂIJAn endâĂŘtoâĂŘend participatory urban noise mapping system,âĂİ. Proceedings of the 9th ACM/IEEE Int Conf on Information Processing in Sensor Networks .New York,USA,ACM,2010:105-116.

[3] Y. Wang, J. Liu, Y. Chen, M. Gruteser, J. Yang, and H. Liu, âĂIJE-eyes: Device-free location-oriented activity identification using fine-grained wifi signatures,âĂİ. Proceedings of the 20th Annual International Conference on Mobile Computing and Networking. New York, USA,2014:617âĂŞ628.

[4] L. Guo, L. Wang, J. Liu, W. Zhou, B. Lu,âĂIJA Novel benchmark on human activity recognition using WiFi signals,âĂİ. Proceedings of 2017 IEEE 19th International Conference on e-Health Networking, Applications and Services.

[5] Q. Pu, S. Gupta, S. Gollakota, and S. Patel, âĂIJWhole-home gesture recognition using wireless signals,âĂİ. Proceedings of the MobiCom, pages 27âĂŞ38. ACM, 2013.

[6] F. Adib, Z. Kabelac, D. Katabi, and R. C. Miller, âĂIJ3d tracking via body radio reflection,âĂİ. Proceedings of the NSDI, pages 317âĂŞ329, 2014.

[7] P. Bahl and V. N. Padmanabhan, âĂIJRadar: An in-building rf-based user location and tracking system,âĂİ. Proceedings of the 19th Annual Joint Conference of the IEEE Computer and Communications Societies. Tel Aviv, Israel, 2000:775âĂŞ784.

[8] S. Sigg,S. Shi ,F. Buesching ,Y. Ji and Wolf, âĂIJL.Leveraging rf-channel fluctuation for activity recognition: Active and passive systems, continuous and rssi-based signal features,âĂİ. Proceedings of International Conference on Advances in Mobile Computing and Multimedia (2013), 43.
[9] S. Sigg, M. Scholz,S. Shi,Y. Ji, andM. Beigl, âĂIJRf-sensing of activities from non-cooperative subjects in device-free recognition systems using ambient and local signals,âĂİ. Mobile Computing, IEEE Transactions on 13, 4 (2014), 907âĂŞ920.
[10] H. Abdelnasser, M. Youssef, and K. A. Harras, âĂIJWigest: A ubiquitous wif-based gesture recognition system,âĂİ. Proceedings of the INFOCOM, pages 1472âĂŞ1480.IEEE, 2015.
[11] G. Wang, Y. Zou, Z. Zhou, K. Wu, and L. M. Ni, âĂIJWe can hear you with wi-fi!âĂİ. Proceedings of the 20th Annual International Conference on Mobile Computing and Networking, New York, USA, 2014:593âĂŞ604.
[12] S. Tan and J. Yang, âĂIJWifinger: Leveraging commodity wifi for fine-grained finger gesture recognition,âĂİ. Proceedings of the 17th ACM International Symposium on Mobile Ad Hoc Networking and Computing, New York, USA, 2016:201âĂŞ210.
[13] K. Ali , Alex X. Liu , and W. Wang. âĂIJKeystroke recognition using wifi signals,âĂİ. Proceedings of ACM MobiCom (2015), 90âĂŞ102.
[14] X. Zheng, J.Wang, L. Shangguan, Z. Zhou, and Y. Liu, âĂIJSmokey: Ubiquitous Smoking Detection with Commercial WiFi Infrastructures,âĂİ. Proceedings of the 35th Annual IEEE International Conference on Computer Communications.2016.
[15] J. Shang and J. Wu, âĂIJA Robust Sign Language Recognition System with Multiple Wi-Fi Devices,âĂİ. Proceedings of ACM SIGCOMM Workshop, Los Angeles, CA, USA, July 2017 (MobiArchâĂŹ2017).
[16] S. Sen, B. Radunovic, R. R. Choudhury and T. Minka, âĂIJYou are Facing the Mona Lisa: Spot Localization using PHY Layer Information,âĂİ. ACM MobiSys, 2012.
[17] X. Wang, L. Gao, and S. Mao, âĂIJCSI Phase Fingerprinting for Indoor Localization With a Deep Learning Approach,âĂİ. IEEE Internet of Things Journal, 2016.
[18] M. Muller, âĂIJDynamic time warping,âĂİ. Information retrieval for music and motion (2007), 69âĂŞ84.
[19] J. C. Platt, âĂIJProbabilistic Outputs for Support Vector Machines and Comparisons to Regularized Likelihood Methods,âĂİ. Cambridge:MIT Press,1999.

# Appendix

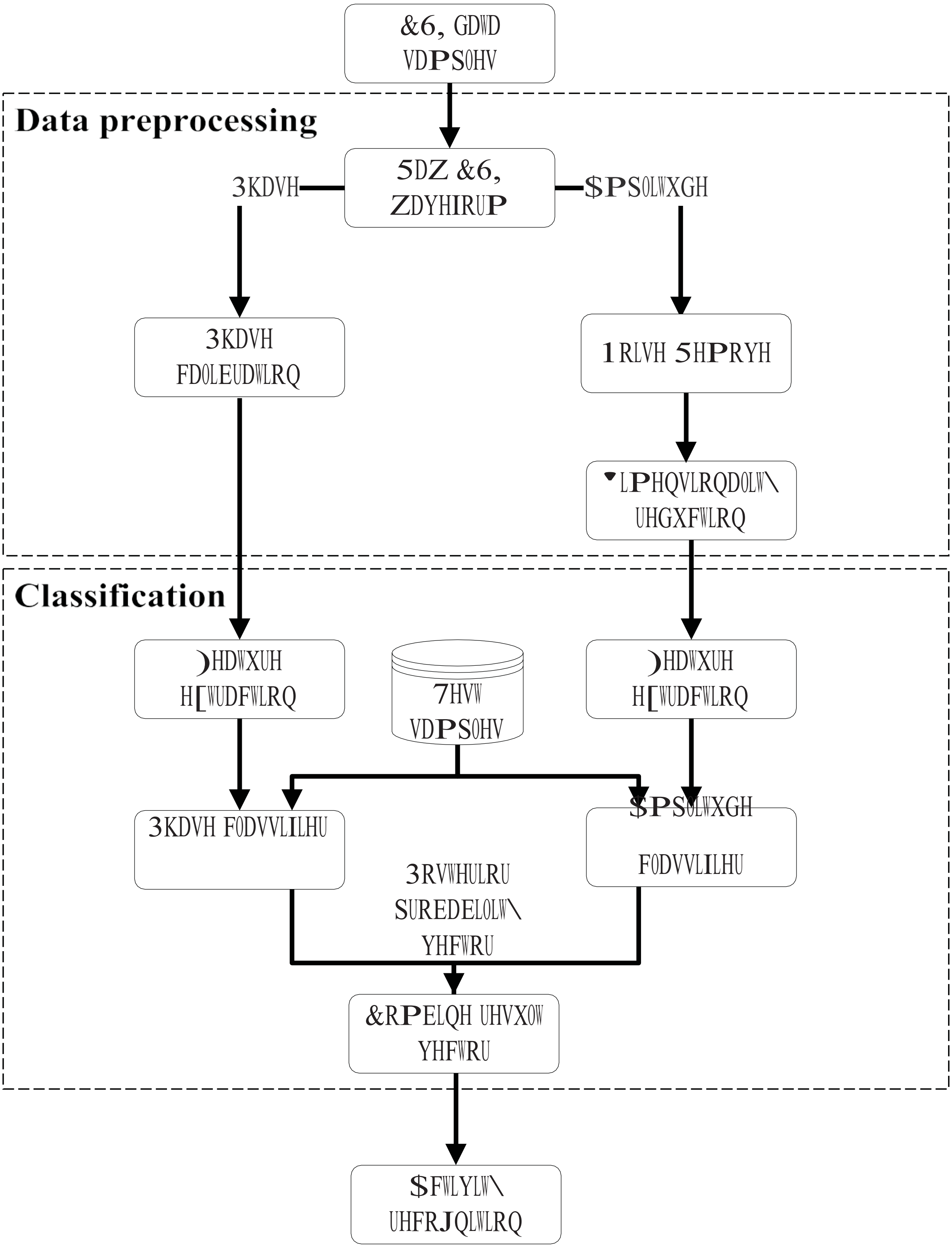

Fig. 2. System structure of Wimotion.

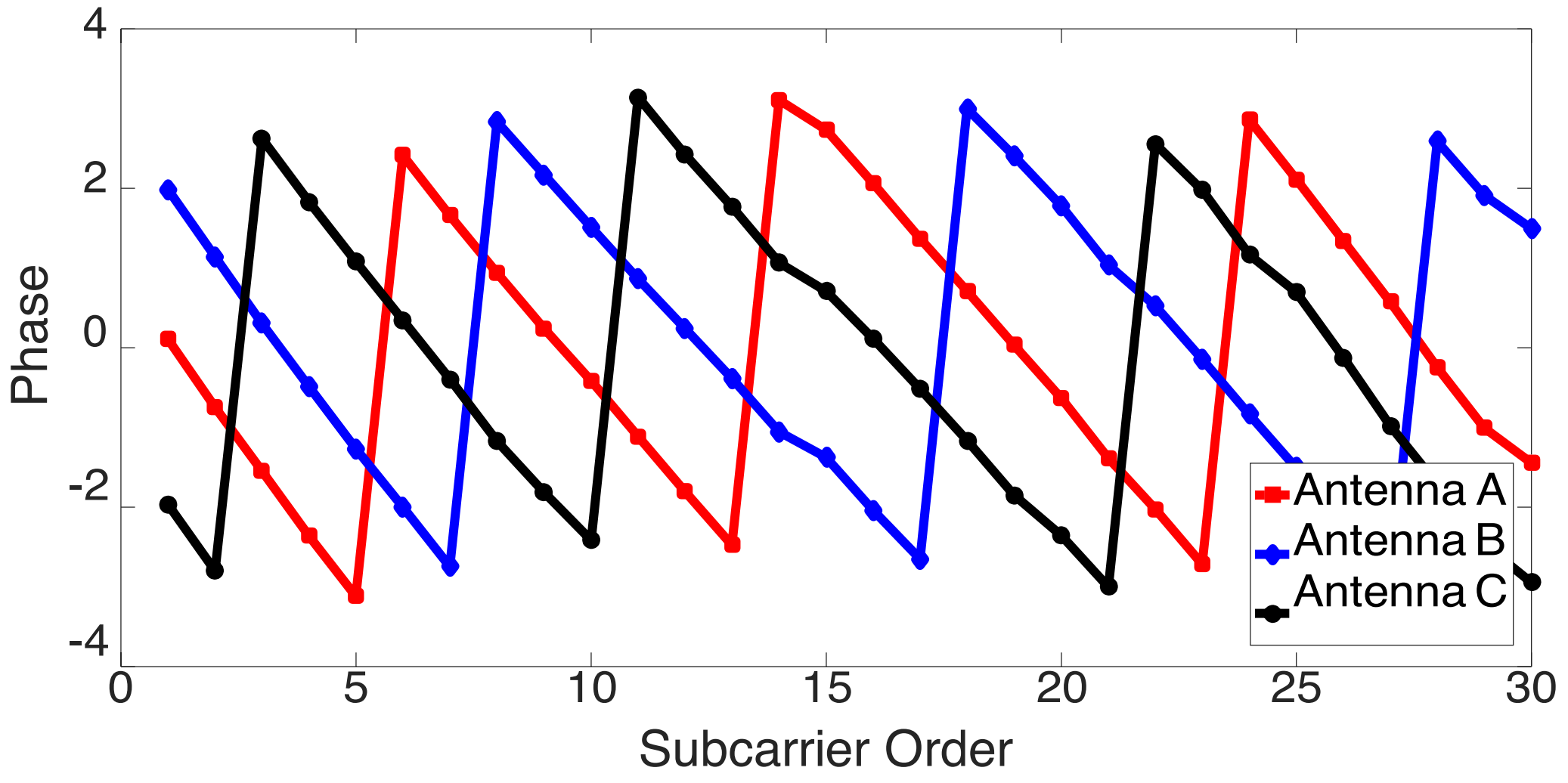

Fig. 3. Raw phase values for three different antennas

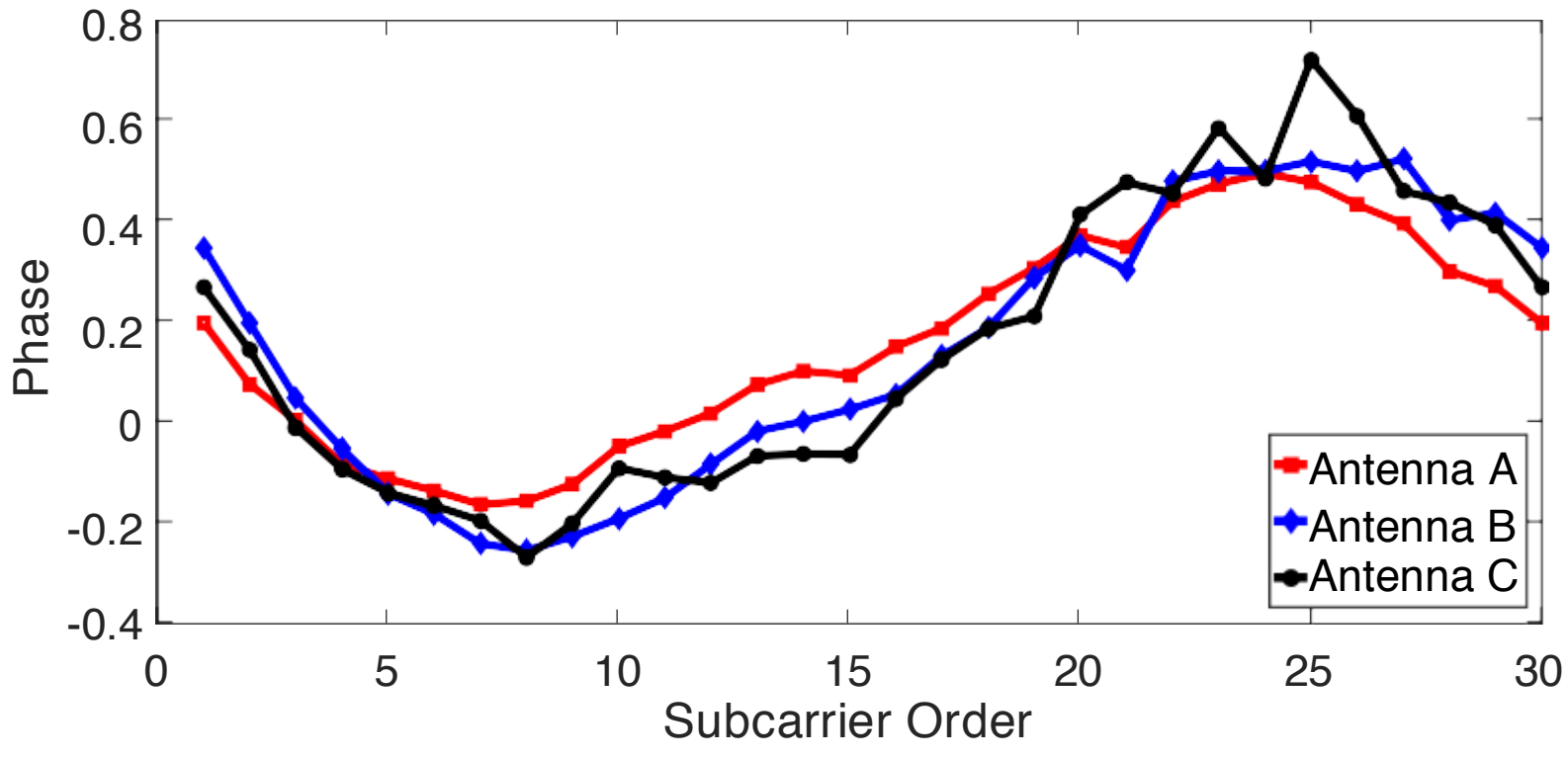

Fig. 4. Transformed phase values for three different antennas.

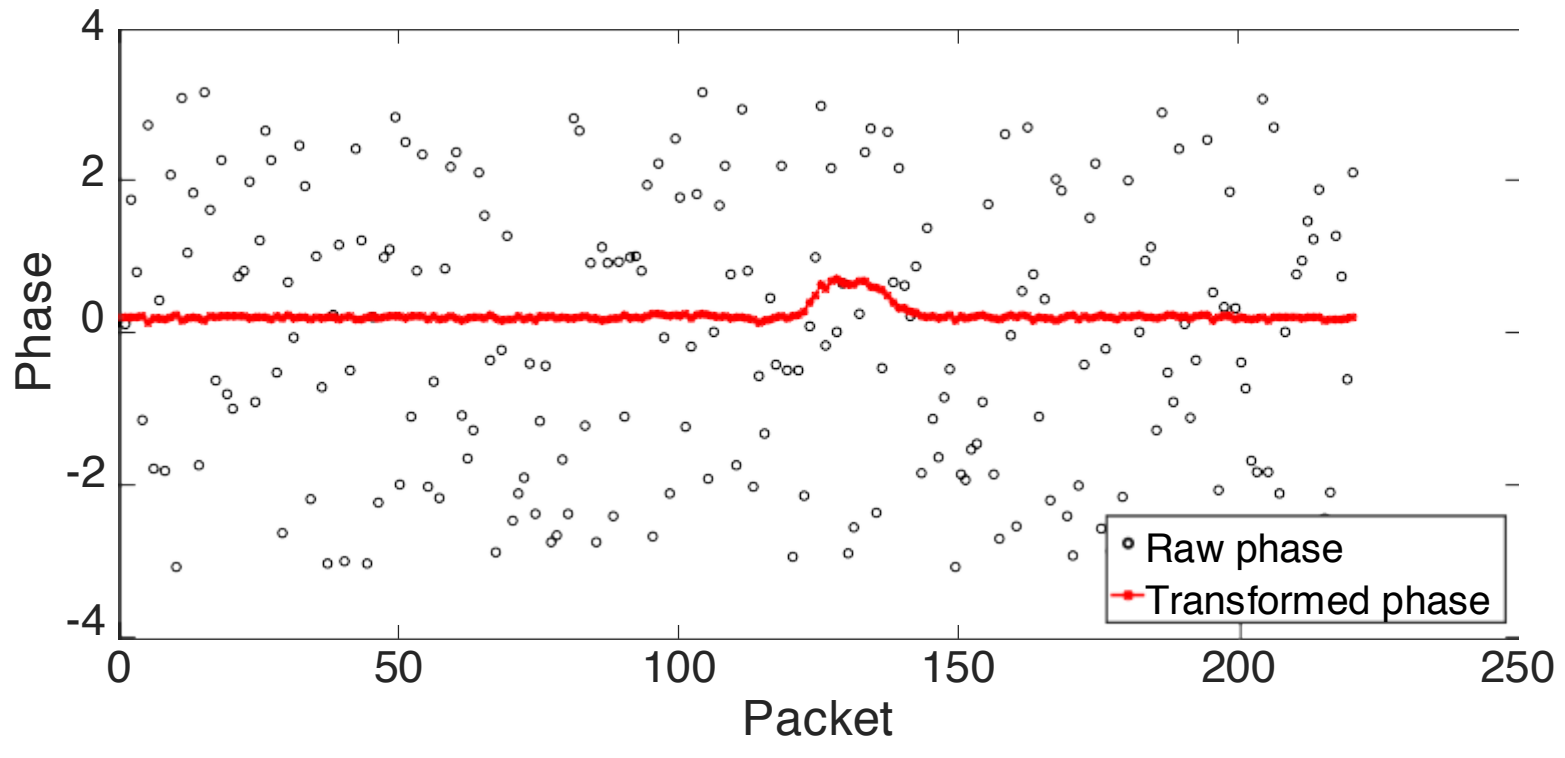

Fig. 5. Random noises are removed in sanitized phase information of a squat sample.

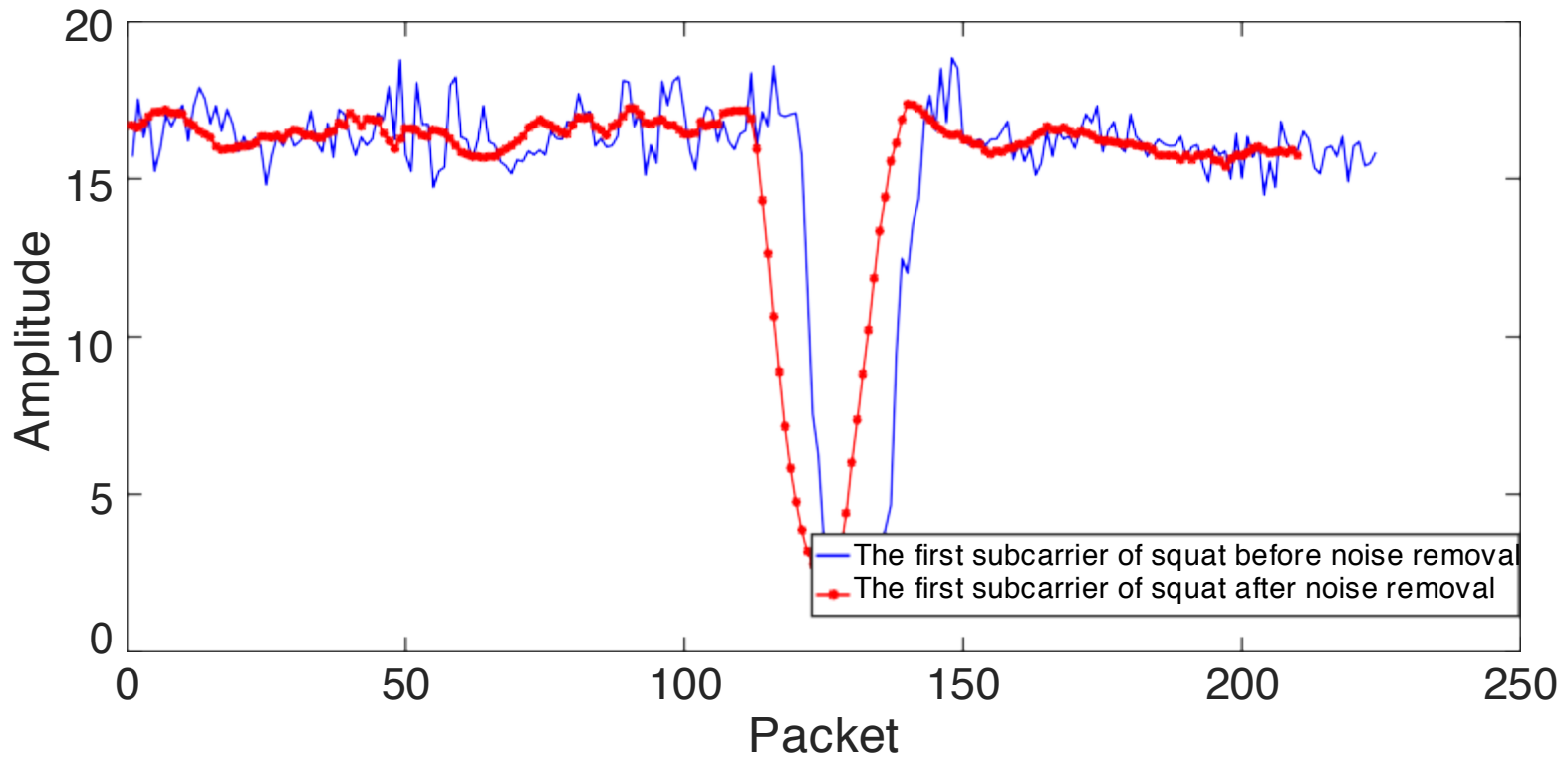

Fig. 6. Weighted moving average.

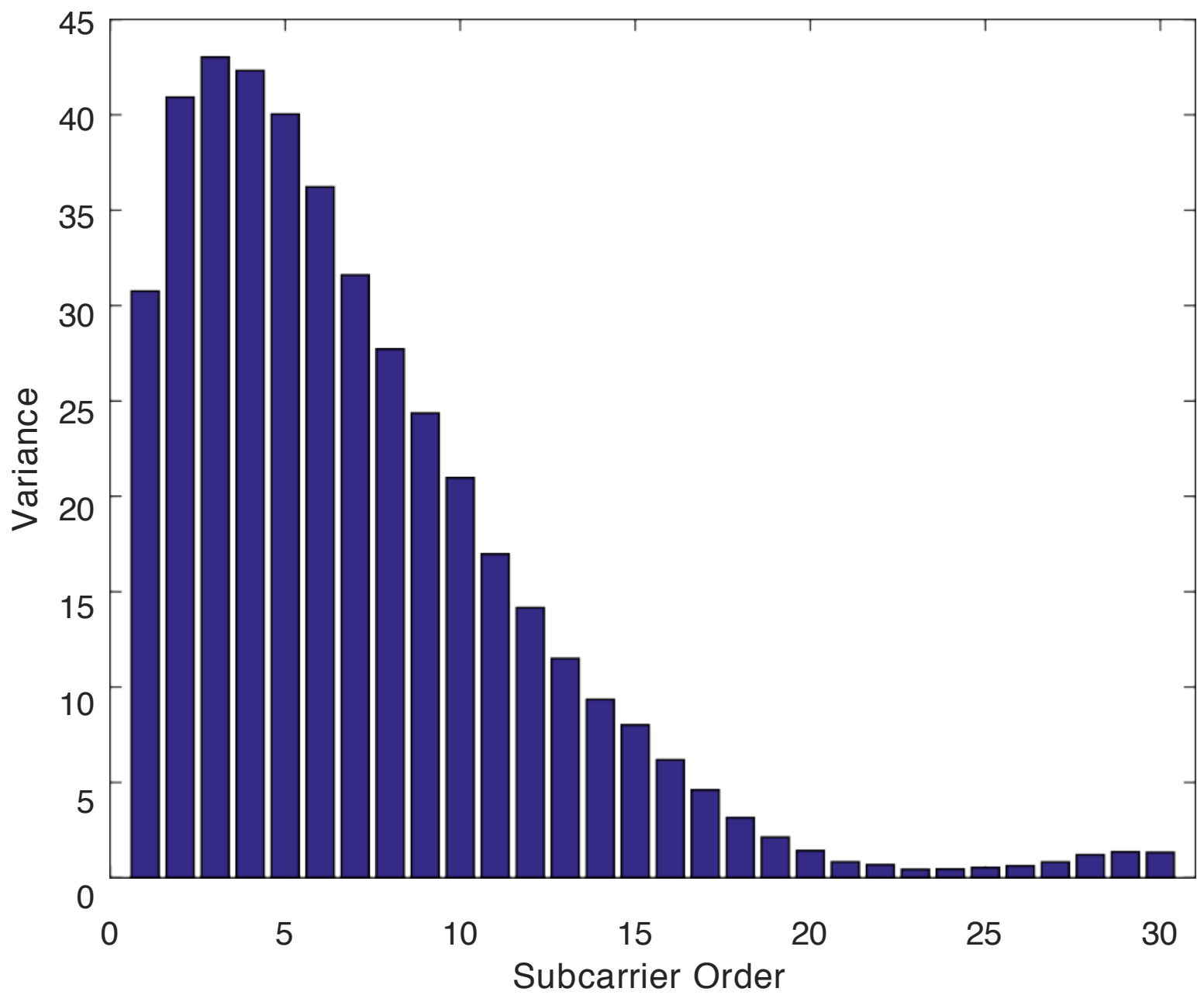

Fig. 7. Different subcarriers have different sensitivities for the same activity.

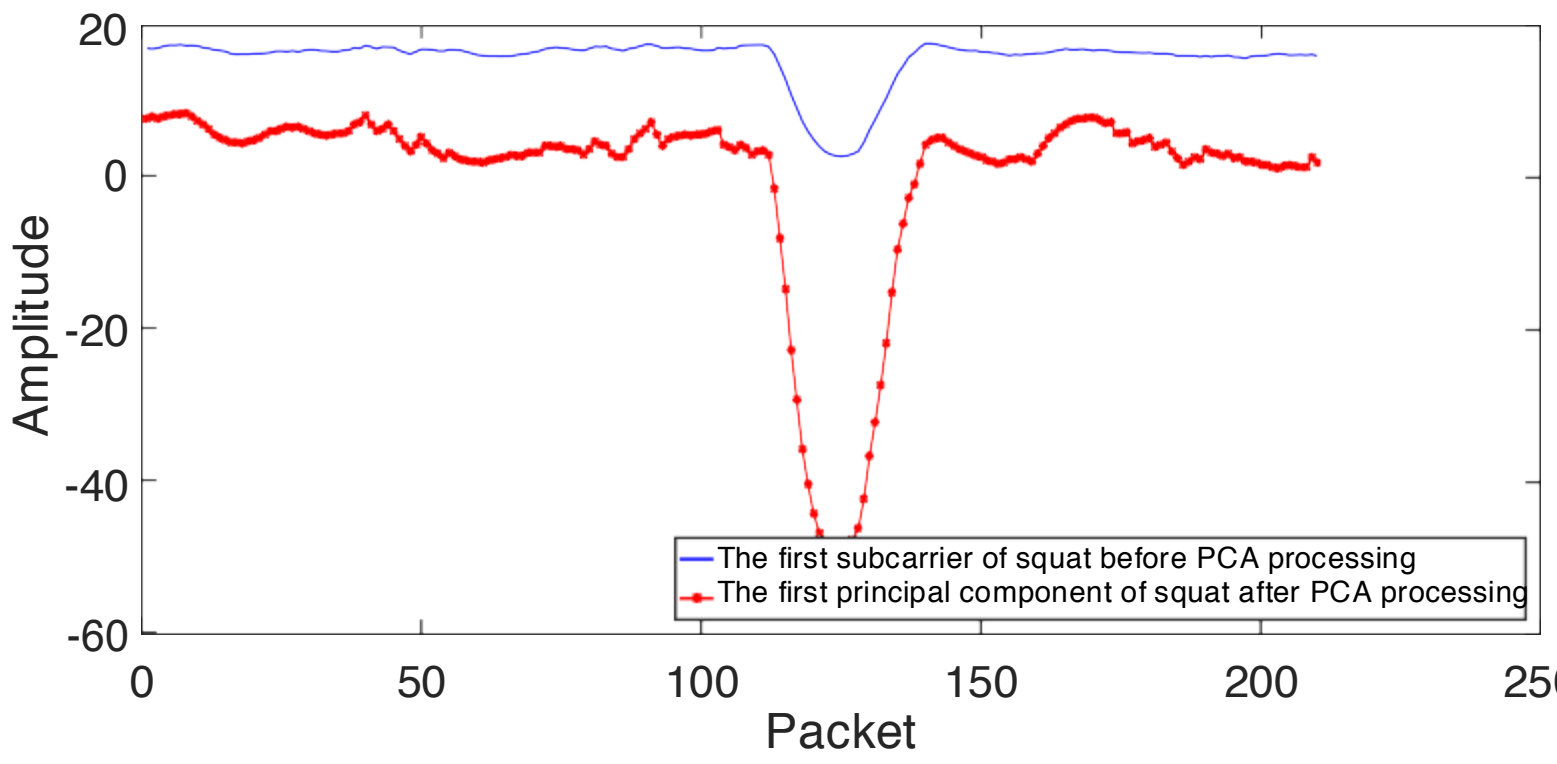

Fig. 8. Principal component analysis.

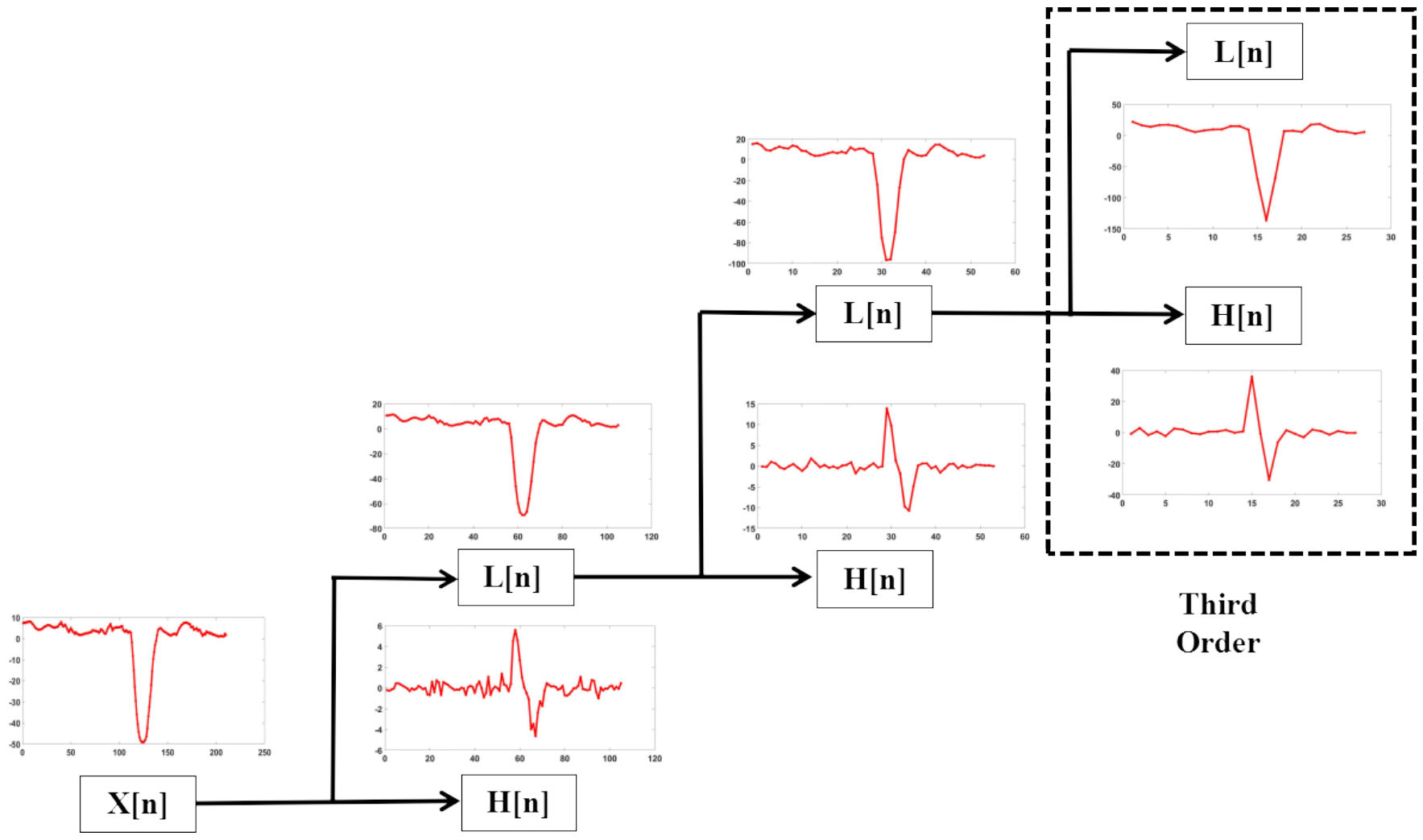

Fig. 9. Discrete wavelet transformation.

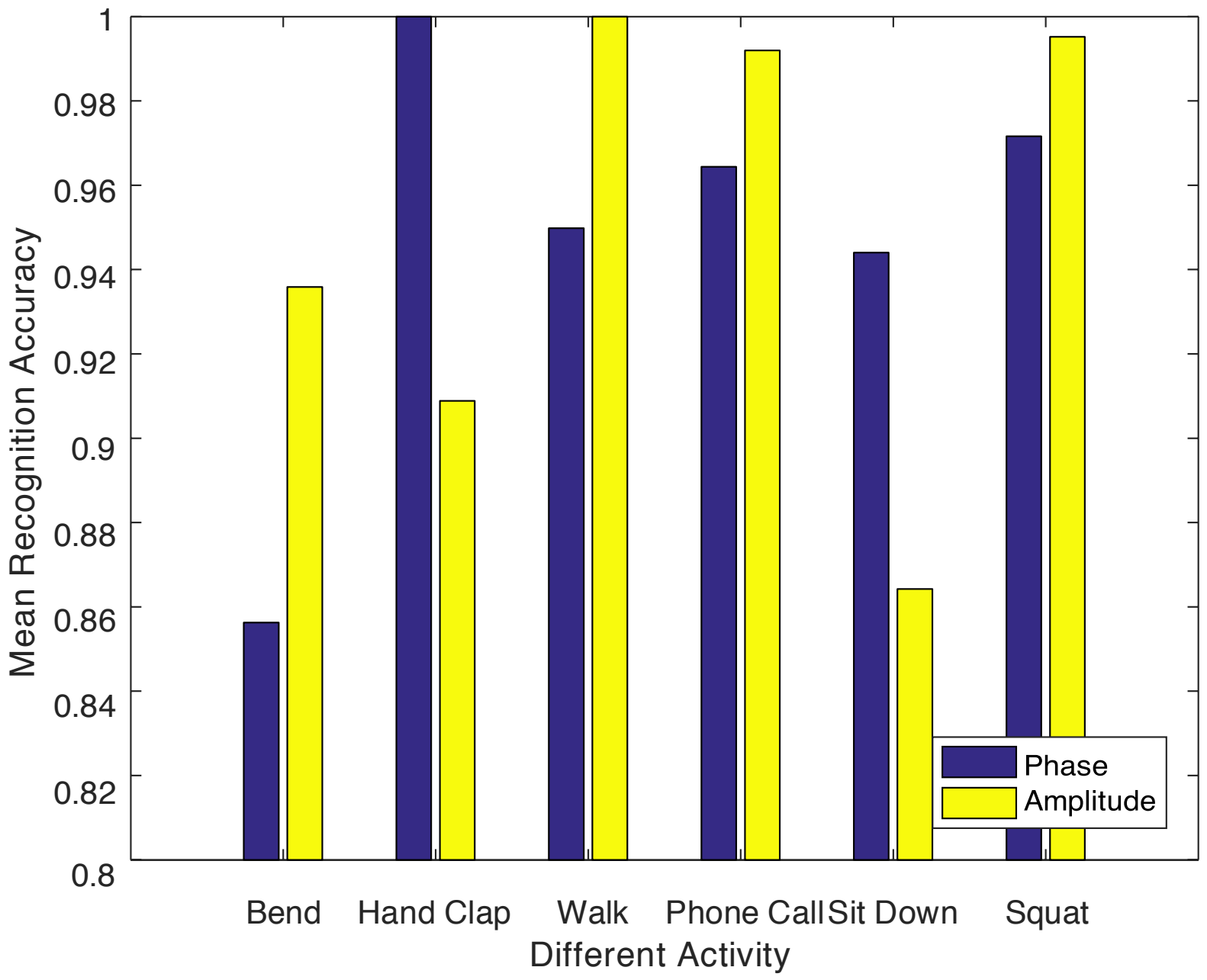

Fig. 10. Classification performance of two classifiers.

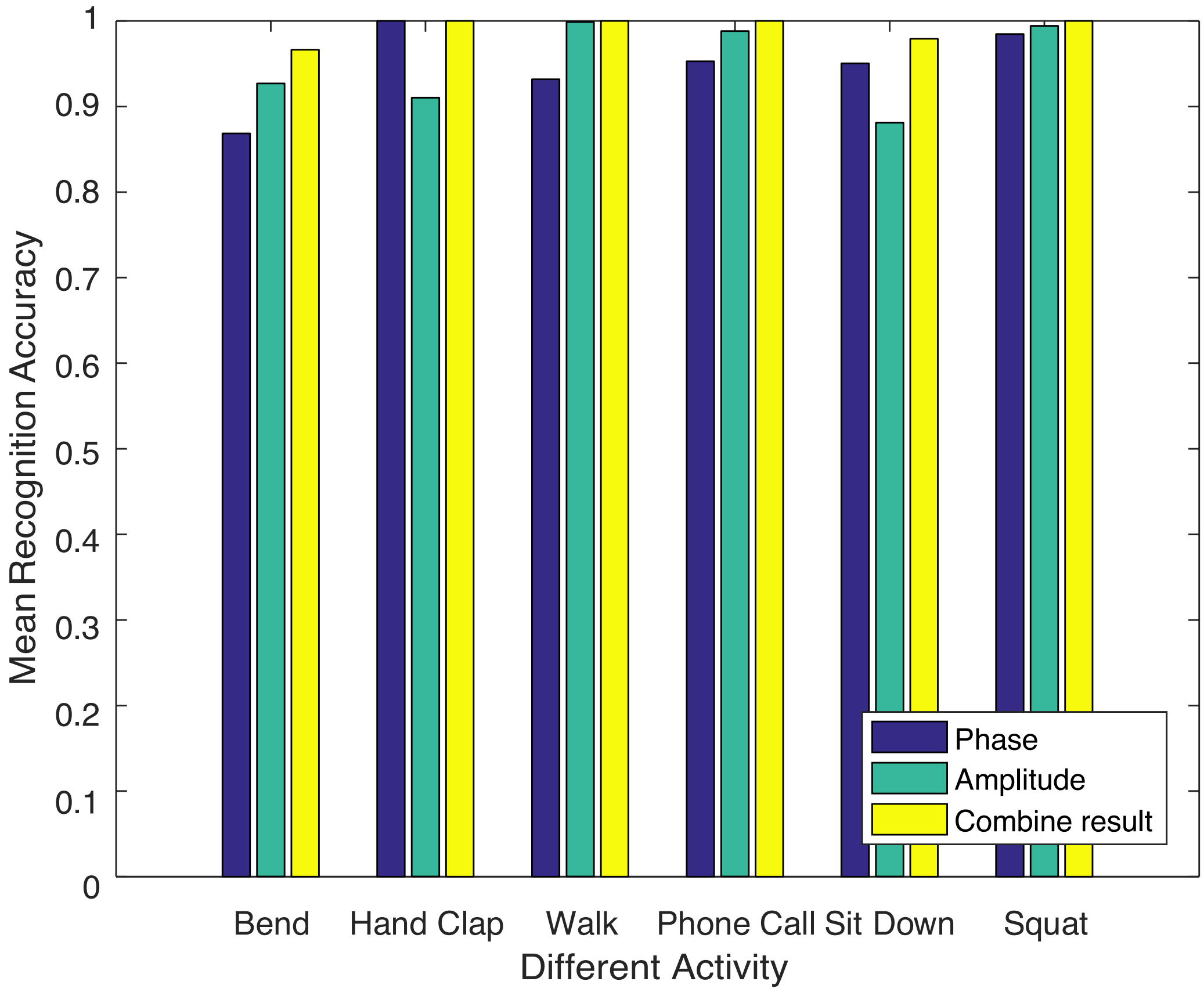

Fig. 11. Prediction accuracy of different activity.

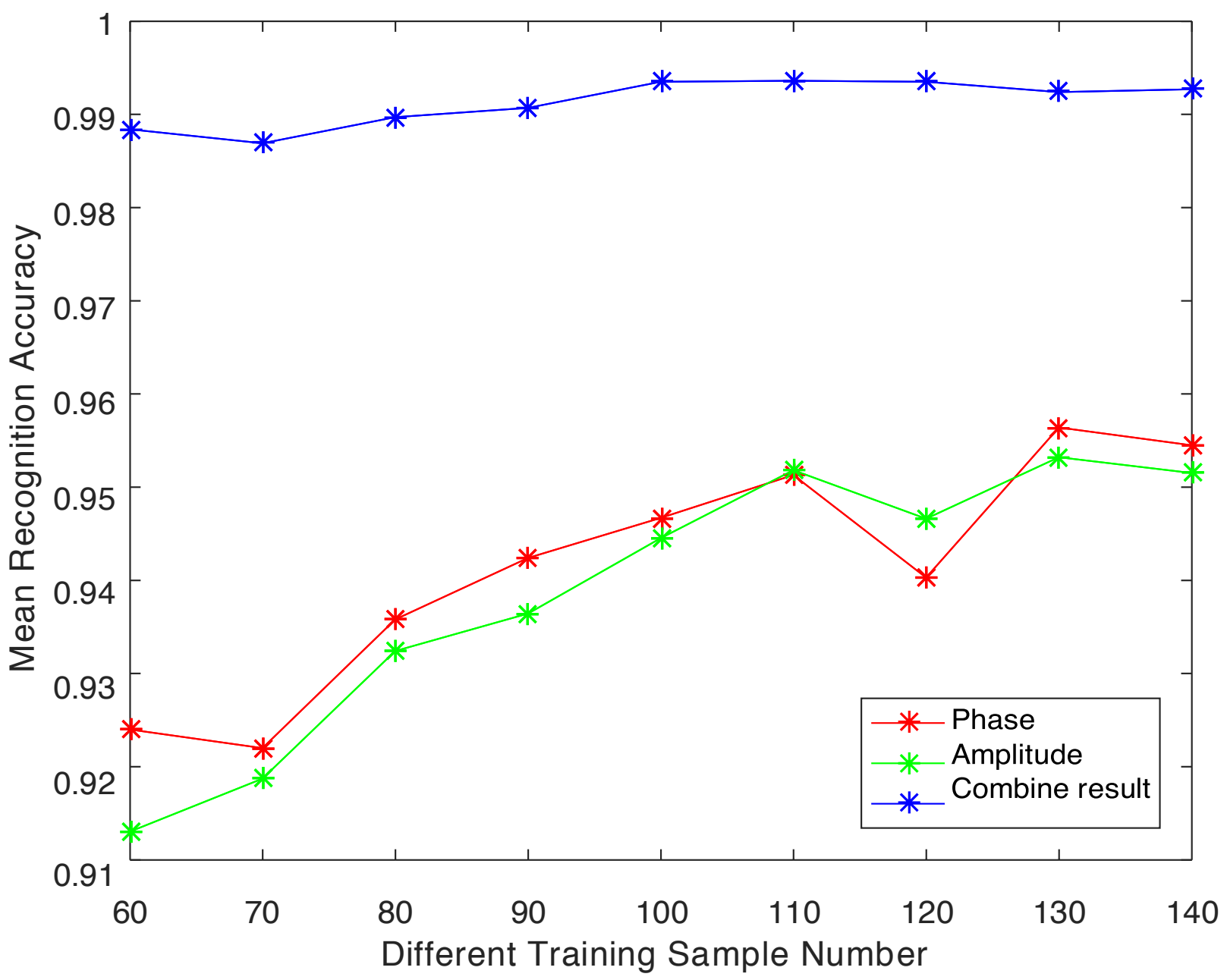

Fig. 12. The influence of different number of training sample.

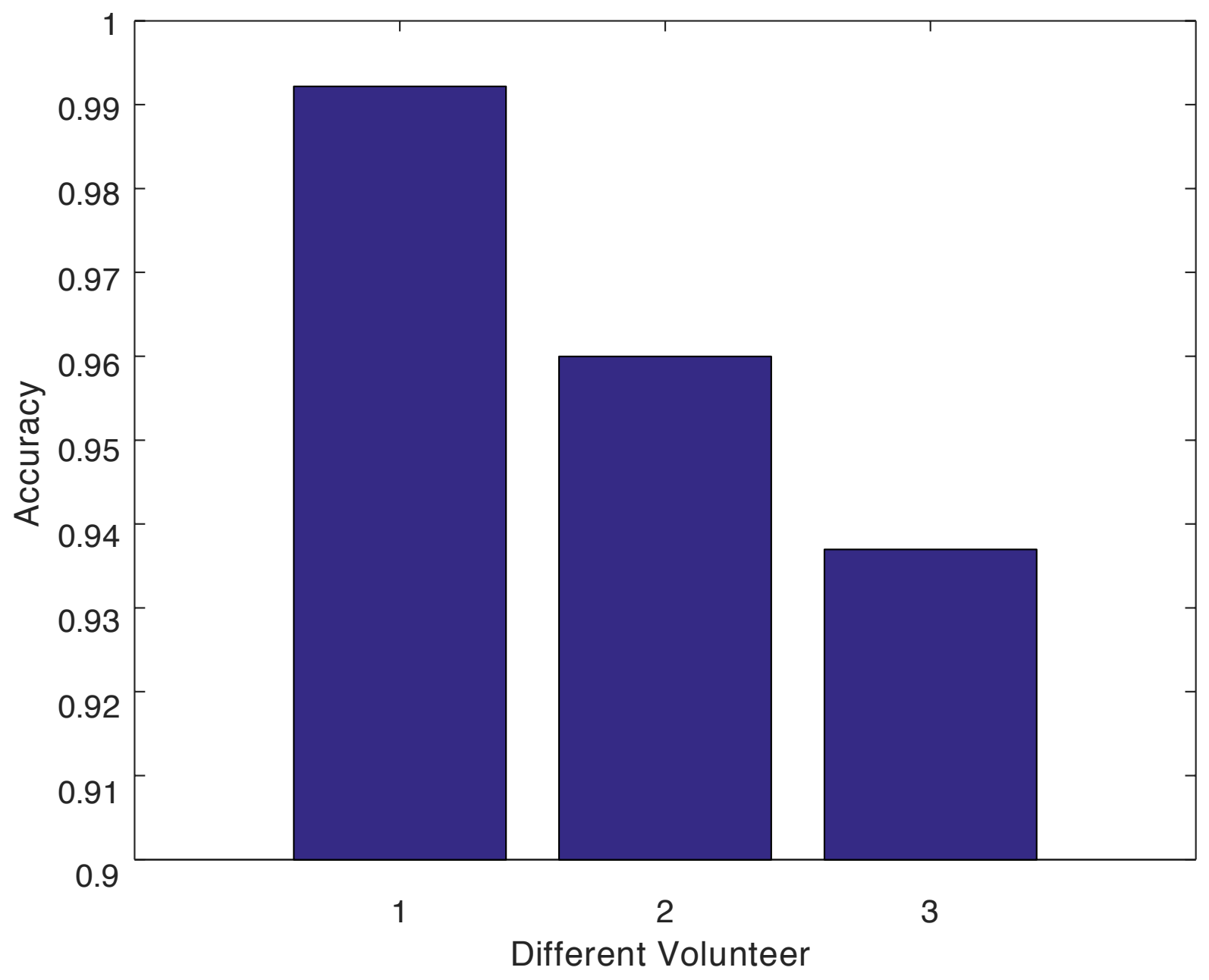

Fig. 13. The influence of different volunteers on combine-result .

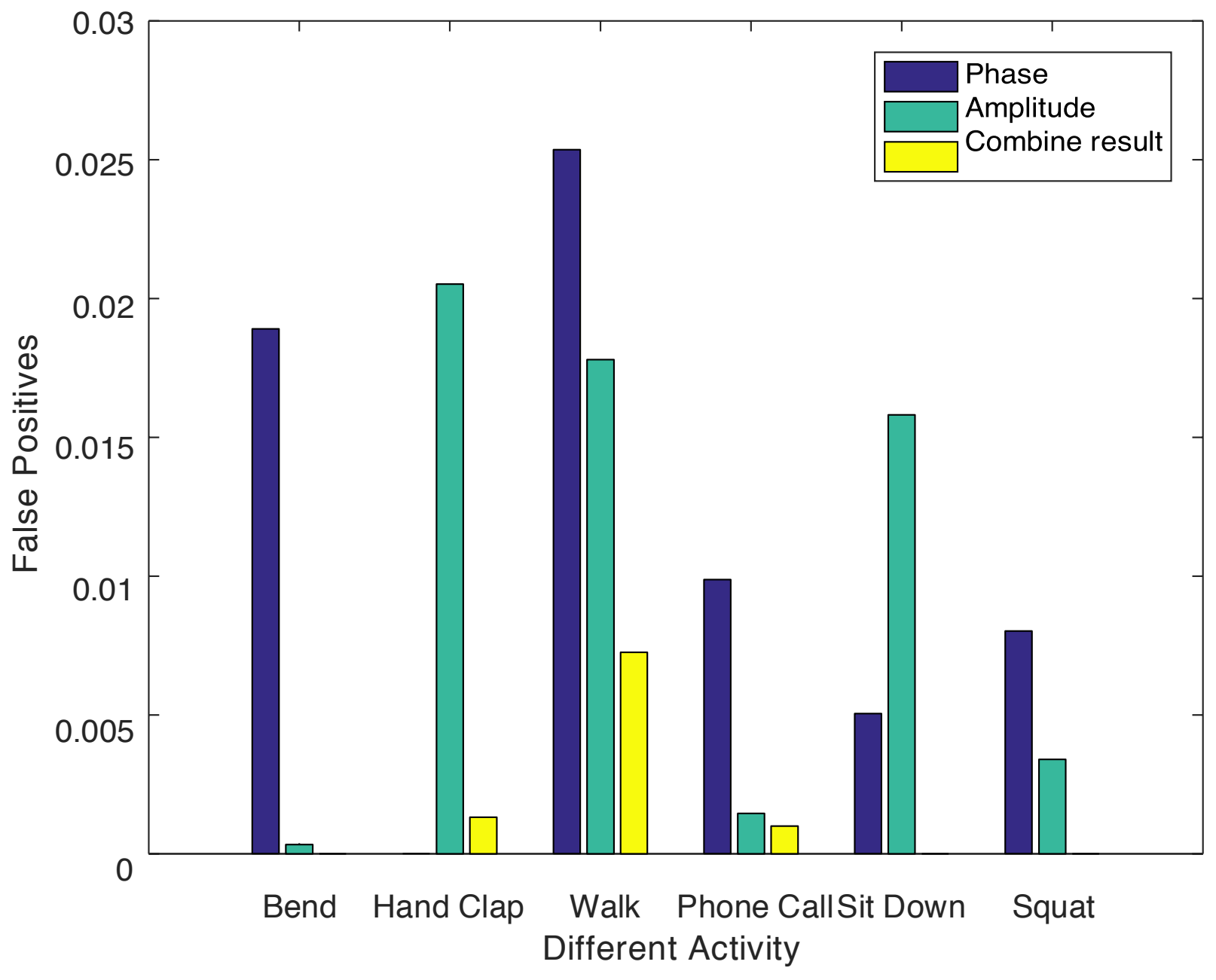

Fig. 14. False positives of different activity.

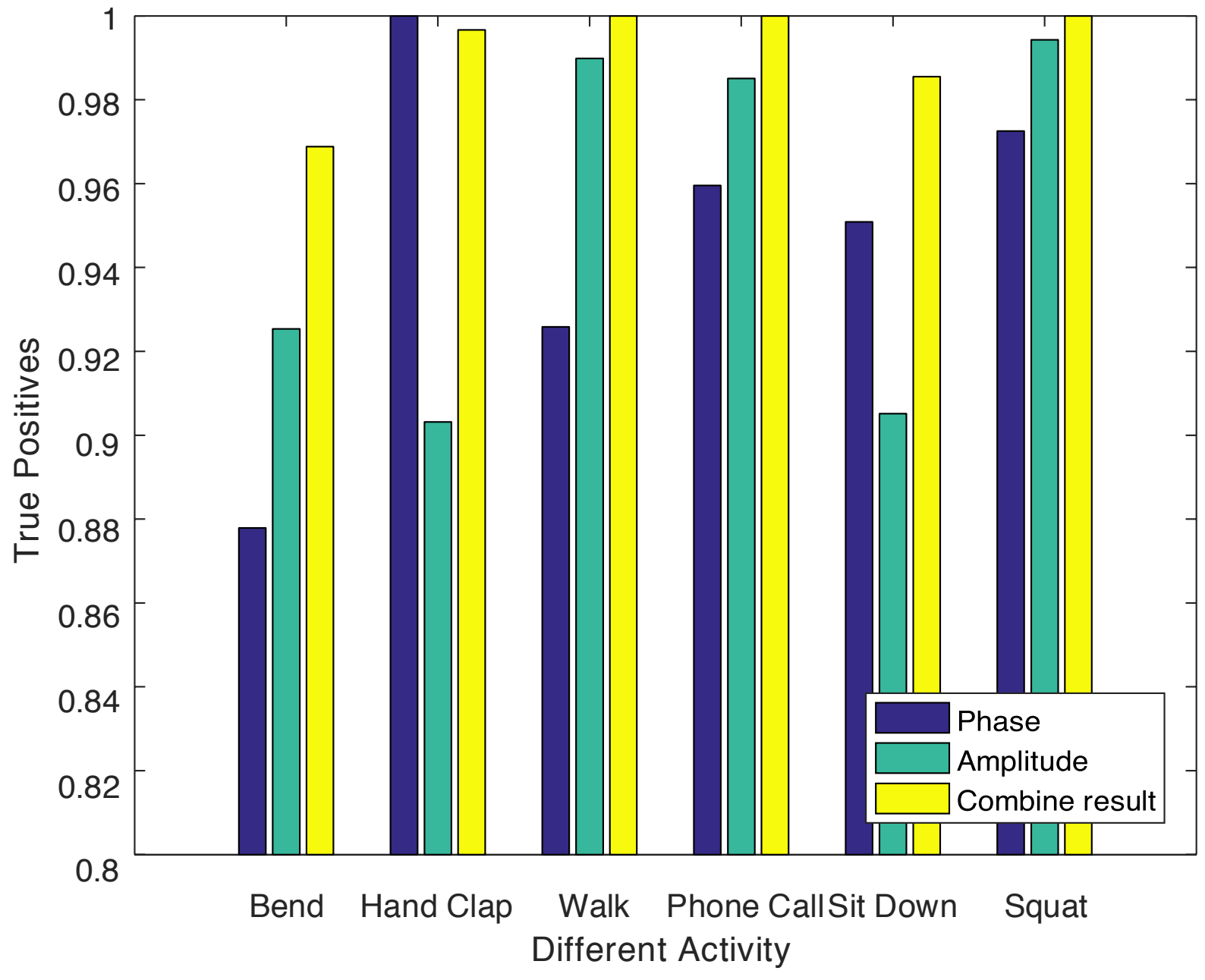

Fig. 15. True positives of different activity.